\documentstyle[aps,epsfig,multicol]{revtex}

\begin{document}

\title{Critical Behaviour in the Spin Fluctuations and Superfluid 
Density of La$_{2-x}$Sr$_x$CuO$_4$}
\author{C. Panagopoulos$^1$, B.D. Rainford$^2$, J.R. Cooper$^1$, 
C.A. Scott$^3$ and T. Xiang$^4$}
\address{$^1$ IRC in Superconductivity and Cavendish Laboratory, 
University of Cambridge, Cambridge CB3 0HE,
United Kingdom}
\address{$^2$ Department of Physics and Astronomy, University of
Southampton, Southampton S017 1BJ, United Kingdom}
\address{$^3$ Rutherford Appleton Laboratory, ISIS Facility, Didcot,
Oxon OX11 0QX, United Kingdom}
\address{$^4$ Institute of Theoretical Physics, Academia Sinica,
P.O. Box 2735, Beijing 100080, Peoples Republic of China}

\date{\today}

\maketitle

\begin{abstract}
We study the doping dependence of low frequency spin fluctuations and the
zero-temperature superfluid density of La$_{2-x}$Sr$_x$CuO$_4$ using the
muon spin relaxation ($\mu$SR) and ac-susceptibility techniques.
Superconductivity is found to coexist with low frequency spin fluctuations
over a large region of the superconducting phase diagram. The characteristic
temperature of spin fluctuations detected by $\mu$SR decreases with
increasing $x$ and vanishes above a critical doping $x_c\sim 0.19$. 
This value of $x_c$ coincides with the doping at
which the normal state pseudogap extrapolates to zero. The superfluid 
density
behaves in the opposite way to the low frequency spin fluctuations. 
It increases with $x$ and becomes nearly doping-independent for $x>x_c$. 
These results are consistent with predictions involving quantum 
criticality at $x_c$.
\end{abstract}

\begin{multicols}{2}

The importance of quantum criticality and the interplay between magnetism and
superconductivity in elucidating the pairing mechanism in high-$T_c$
superconductors (HTS) have been discussed in many theoretical 
and experimental works  since their discovery \cite
{SC,CMV,SY,CC,CV1,SCRL,DP,Harshman,Weid,Kiefl,Budnick,Berthier,GA1,KY,CN,HA}. 
Spin fluctuations and quantum criticality have been extensively 
studied in heavy fermion compounds, where the appearance of superconductivity 
as the Neel temperature is suppressed
with pressure, has provided strong evidence both for a quantum critical 
point and for a pairing mechanism mediated by spin fluctuations
\cite{Mathur}.  
Unlike heavy fermion compounds the properties of HTS are usually tuned 
by varying the carrier concentration (doping). Fundamental physical quantities
such as the superconducting condensation energy \cite{JWL} and the superfluid 
density  \cite{LSC} show distinct changes at a critical doping $x_c$ 
which is slightly above the optimal carrier concentration defined as the point 
at which the superconducting transition temperature, $T_c$, is maximum. 
Recently, Tallon and Loram \cite{TallonLoram} have analysed the doping dependence 
of the normal state
pseudogap from the specific heat and a set of other experimental data and
suggested that the normal state pseudogap vanishes abruptly at $x_c$, as 
predicted for quantum criticality  \cite{SC,CMV,SY,CC,CV1,SCRL}. 
It is therefore important to search for a critical behaviour in HTS 
using direct measurements of the spin or charge fluctuations. 

In this Letter we present a systematic study of the doping dependence of
low frequency spin fluctuations and the superfluid density of La$_{2-x}$Sr$_x
$CuO$_4$. From the muon depolarisation and the in-plane penetration
depth determined both by $\mu $SR and ac-susceptibility, we identify a critical
doping $x_c$ at which qualitative changes occur in both the
spin  excitations and the superfluid density. For $x<x_c$ there are 
significant low frequency magnetic fluctuations whereas the superfluid 
density changes in the opposite manner.
Above $x_c$ these fluctuations are not detected and the superfluid density
becomes nearly doping independent as expected for conventional BCS superconductors. 
The correlation between these two quantities provides evidence for 
the presence of a quantum critical point, in support of theoretical 
predictions made by several groups \cite{SC,CMV,SY,CC,CV1,SCRL,DP}.

Zero-field $\mu $SR is a sensitive modern technique for studying  
low frequency spin dynamics. 
It is a local probe and can detect very small (1$G$) internal
magnetic fields \cite{YU}. The data obtained by $\mu $SR give a 
direct measure of the low energy spin dynamics in the
sample \cite{YU}. From the time dependence of the depolarisation of muons, one can
obtain important information about the characteristic energy or
temperature scale of spin fluctuations and the spin-glass transition
temperature. 

La$_{2-x}$Sr$_x$CuO$_4$ (LSCO) is one of the simplest HTS systems. 
It has a simple crystal structure
and its carrier concentration can be accurately controlled. 
Undoped La$_2$CuO$_4$ is an insulator with long range antiferromagnetic
order. Substitution of Sr$^{2+}$ for La$^{3+}$ introduces holes into the 
CuO$_2$ planes and at $x\simeq 0.02$ the system exhibits a short range ordered,
antiferromagnetic or spin glass phase \cite{KY,CN}. Superconductivity
emerges near $x=0.05$ and the superconducting transition temperature approximately follows
a parabolic doping dependence and vanishes at $x\sim 0.30$ \cite{Torrance}.

The samples we studied were high quality polycrystalline 
La$_{2-x}$Sr$_x$CuO$_4$ ($x=0.03-0.24$) synthesised in Cambridge. 
The phase purity of these samples,
as characterised by powder x-ray diffraction and micro-Raman spectroscopy \cite{EL} as
well as extensive transport \cite{JRC} and thermodynamic 
measurements \cite{JWL,Loram,CP4} was found to be
better than $1\%$. Their lattice parameters and $T_c$ values
were also in good agreement with published data obtained on powders as well
as single crystals \cite{PGR,KY}.

Zero-field (ZF) and transverse-field (TF) $\mu $SR experiments were
performed at the pulsed muon source, ISIS Facility, Rutherford Appleton
Laboratory and spectra were collected over the temperature range from $40mK$
to $150K$. In a $\mu $SR experiment, $100\%$ spin polarised positive muons
are implanted into a specimen and evolve in their local magnetic environment.
The muon decays with a life time $2.2\mu s$, emitting a
positron preferentially in the direction of the muon spin at the time of
decay. By accumulating time histograms of such positrons one may deduce the
muon depolarisation rate as a function of time after implantation. The
pulsed muon facility at ISIS allows the muon polarisation function to be 
measured for up to 6-7 muon life times,
and thus small changes in the depolarisation function can be detected. 
This property of the ISIS muon facility has allowed us to 
study the temperature dependence of slow
spin fluctuations in a HTS. The muon is expected to reside in the most
electronegative site of the lattice. In the case of LSCO this is 1.0\AA\ away
from the apical O$^{-2}$ site of the CuO$_2$ plane \cite{muonsite}. 
As will be discussed later, this assignment is supported by the strong
enhancement in the muon depolarisation rate with Zn doping since Zn is known
to substitute in the CuO$_2$ planes. We therefore believe that the results
reported here are dominated by the magnetic correlations in the CuO$_2$ planes.

\begin{figure}
\begin{center}
\epsfig{file=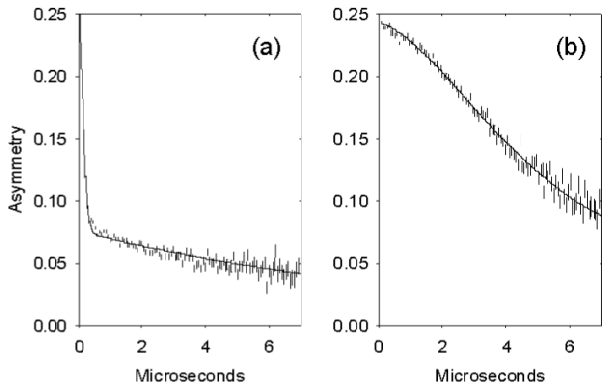,width=5.5cm,clip=,angle=-90}
\end{center}
\caption{Typical zero-field $\mu $SR spectra of La$_{2-x}$Sr$_x$CuO$_4$ for 
$x=0.08$ measured at (a) $1.3K$ and (b) $9K$. The solid line in (b) is a fit to a
stretched exponential, discussed in the text. 
In (a) the solid line is a fit to a Lorentzian, decribing the initial rapid drop of 
asymmetry, plus a stretched exponential, describing the long-time tail. 
}
\label{fig1}
\end{figure}

The zero temperature superfluid density, $\rho ^s(0)$, was determined from
measurements of the magnetic penetration depth $\lambda _{ab}$ for current 
flow in the CuO$_2$ planes, ($\rho ^s$$\sim $$\lambda _{ab}^{-2}$). $\lambda _{ab}$ was
obtained for all samples using the low-field $ac$ -susceptibility technique
(typically at $1G$ and $333Hz$) for grain-aligned powders \cite{CP1}. Details
of the technique can be found in refs \cite{CP4,CP1}. For each doping
concentration of LSCO, the result presented here is typical of those
obtained for 2-4 samples prepared independently by aligning powders from the same
polycrystalline pellet. Four pieces were cut out from each aligned sample and measured. 
In other words 8-16
samples were investigated for each doping content. The values of $\lambda
_{ab}^{-2}(0)$ for samples with $x$$\geq 0.15$ were also confirmed by
standard TF-cooled $\mu $SR experiments performed on unaligned powders at
400 Gauss. In a TF-cooled $\mu $SR experiment, the field
distribution of a flux-line lattice produced by an external field is probed
by muons. The depolarisation rate of the initial muon spin is proportional
to $\lambda _{ab}^{-2}(0)$\cite{CP2,CP3}. For $x<0.15$, $\lambda
_{ab}^{-2}(0)$ obtained from TF-$\mu $SR is less accurate than that obtained
from the $ac$-susceptibility measurements because of a strong
enhancement in the muon depolarisation rate below $0.3$ $T_c$  \cite{CP3} probably 
due to the rapid increase in slow spin fluctuations with underdoping as discussed below.

\begin{figure}
\epsfig{file=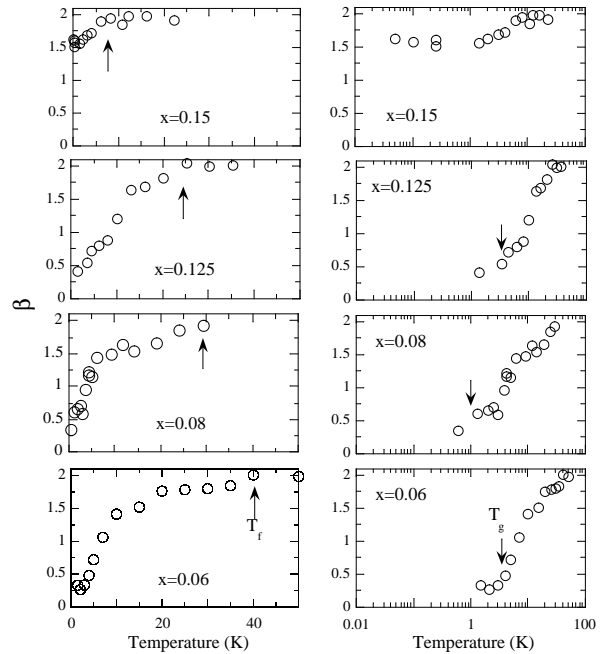,clip=,width=8cm,angle=0}
\caption{Temperature dependence of the exponent $\beta$ of 
La$_{2-x}$Sr$_x$CuO$_4$ for x=0.06, 0.08, 0.125, 0.15. The left hand panel 
shows linear plots with the arrows indicating $T_f$ whereas in the right 
hand panel we show semi-log plots with the arrows indicating $T_g$. 
The error in $\beta$ is $\pm 0.06$. 
}
\end{figure}

Figure 1 shows the typical time dependence of the ZF muon asymmetry 
for $x=0.08$ ($T_c=21K$) for two characteristic temperatures. In all samples the
high temperature form of the depolarisation is Gaussian, consistent with
dipolar interactions between the muons and their near neighbour nuclear
moments. This was verified by applying a $50G$ longitudinal field, which
completely suppressed the depolarisation. The onset of dynamical relaxation
processes becomes apparent at low temperatures 
by a deviation of the depolarisation from a Gaussian behaviour. 

The depolarisation rate can be well described
by a stretched exponential function $G_z(t)=A_1
exp (- (\gamma t)^\beta) +A_2$, where $\gamma$ is the
dynamic muon spin depolarisation rate and $A_2$ 
accounts for a small time independent
background arising from muons stopping in the
silver backing plate.  As can be seen in Fig. 2
the exponent $\beta$ starts from the value 2 at
high temperatures, corresponding to Gaussian
relaxation, but decreases steadily at lower
temperatures, suggesting the development of a
low frequency component in the spectral weight
of the spin dynamics.   We take the temperature
at which $\beta$ first drops below the value 2 as
one indicator of the energy scale of the
magnetic correlations in a given sample.  This
temperature is denoted $T_f$.  At low temperatures
(Fig. 2) the values of $\beta$ decrease towards the
value 0.5.  At lower temperatures still the form
of the relaxation function changes (Fig. 1a):
there is an initial rapid drop of asymmetry,
followed by a long-time tail with a slower
relaxation.  This is very charactersitic of the
behaviour found in spin glass samples below $T_g$
\cite{YU}, where the initial rapid drop is ascribed
to the static distribution of random local
fields while the long-time behaviour results
from dynamical processes.  In this regime the
data were fitted to the form $G_z(t)=A_1 exp( -
\gamma_1 t) +A_2 exp(- (\gamma_2 t)^\beta) +A_3$.  We
identify the spin glass temperature $T_g$ as the
temperature at which the value of $\beta$ (Fig. 2)
reaches the value 0.5.  This "root exponential"
form for the relaxation function is a common
feature of spin glasses \cite{YU,BDR}, and in the
present samples this temperature coincided with
the cross-over in behaviour (as shown in Fig. 1)
and with a peak in the longitudinal relaxation
rate.  All the samples with $x=0.03-0.125$
followed the same behaviour.

Our data indicate that the spin glass phase persists beyond $x=0.125$. In fact
the onset of the spin glass phase for $x=0.125$ occurs at a higher
temperature than that for $x=0.10$. This may be due to the
formation of strongly correlated antiferromagnetic stripe domains in this
range of doping \cite{KY,CN,Kiv}. For $x=0.15$ and $0.17$, $T_g$ becomes very
small ($<45mK$) and $T_f$ is approximately equal to $8K$ and $2K$,
respectively. For $x\geq 0.20$, there are no changes in the depolarisation 
function to the lowest temperature measured ($40mK$). The presence of a finite 
$T_g$ for $x>0.125$ is verified by the Zn doping studies discussed below.

Figure 3 shows the doping dependences of $T_g$ and $T_f$ together with $T_c$. 
Although the freezing of spins
occurs at very low temperatures, low frequency spin fluctuations appear at
significantly higher temperatures. Both $T_g$ and $T_f$ are found to decrease 
with increasing doping and tend to zero at $x_c\simeq 0.19$. Their behaviour 
resembles that of the normal state gap \cite{JWL,TallonLoram,JRC} determined 
from measurements above $T_c$, which also tends to vanish at the same $x_c$. This is
better demonstrated in the semi-log plot of $T_g$ and $T_f$ shown in the
inset of the figure. The similarity in the doping dependence of $T_g$ and $T_f$ 
suggests that both quantities may have the same scaling behaviour. The
exponential doping dependence of $T_g$ has also been found in 
Y$_{1-y}$Ca$_y$Ba$_2$Cu$_3$O$_{6.02}$ for doping of up to 0.09 holes
per planar Cu atom \cite{CN}. This suggests that the trend of $T_g$ shown in
Fig. 3 is common to all high-$T_c$ materials.

The present $\mu$SR data strongly suggest that there is a certain carrier 
concentration at which two magnetic energy scales associated with the 
superconducting state vanish. Although they are both much smaller than 
the normal state gap, there is a similarity in that they all decrease with 
doping and extrapolate to zero at $x_c\simeq 0.19$ \cite{JWL,TallonLoram,JRC} 
suggestive of a quantum critical point  \cite{CV1,SCRL}.

To further understand the physical properties near $x_c$ we have performed 
detail measurements of the absolute values of the in-plane penetration depth 
for samples prepared from the same batches as those measured by ZF-$\mu$SR. 
The lower panel of Fig. 3 shows the doping dependence of $\lambda_{ab}^{-2}(0)$. 
In contrast to $T_f$ and $T_g$, $\lambda _{ab}^{-2}(0)$ increases
with doping and becomes nearly doping independent above $x_c$. We note the 
inverse relationship between $\lambda _{ab}^{-2}(0)$ and the magnetic effects 
determined by $\mu$SR. The suppression of $\lambda_{ab}^{-2}$ in the underdoped 
region has been reported from the early days
of HTS \cite{CP2}. Recent penetration depth measurements in LSCO and 
HgBa$_2$CuO$_{4+x}$ showed that not only the in-plane but also $c$-axis zero 
temperature superfluid density tend to saturate above $x_c$ \cite{LSC,CP4}. 
The present results allow us to identify a correlation between 
$\lambda _{ab}^{-2}(0)$ and the magnetic effects determining $T_g$ and $T_f$.

\begin{figure}
\begin{center}
\epsfig{file=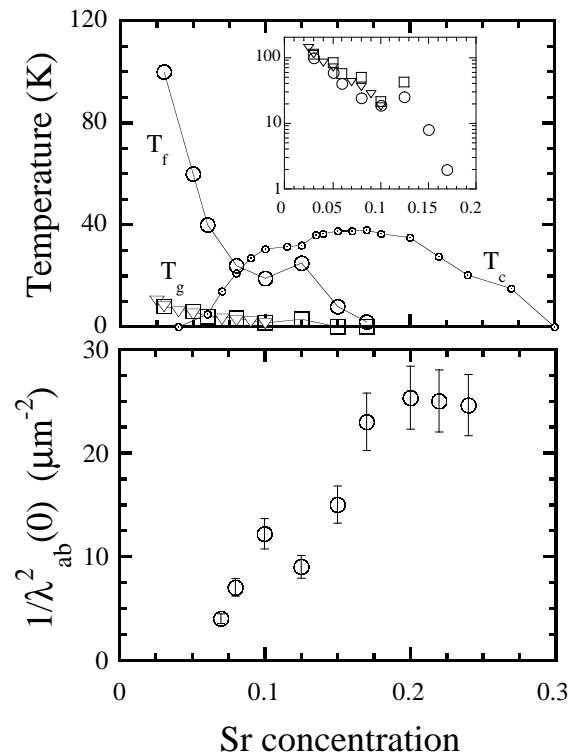,width=8cm,clip=,angle=0}
\end{center}
\caption{
Doping dependence of $T_g$ (squares), $T_f$ (circles), 
and $T_c$ (dotted circles) of La$_{2-x}$Sr$_x$CuO$_4$. $T_g$ data from
Ref. [15] (triangles) are also shown for comparison. The inset is a
semi-log plot of $T_g$ (multiplied by 14) and $T_f$ as a function of doping.
The lower panel shows the doping dependence of the inverse square of the
zero temperature in-plane penetration depth
measured by the $ac$-susceptibility technique.
}
\end{figure}

Distinct changes at $x_c$ are not limited to the properties presented here, 
but have also been found in many other quantities of HTS. As mentioned already, 
the superconducting condensation energy and specific heat jump are maximum at 
$x_c$ and drop quickly with underdoping \cite{JWL,TallonLoram,JRC} furthermore 
the temperature dependence of $\lambda _{ab}^{-2}$ only obeys the $d$-wave 
weak-coupling BCS formula for $x>x_c$ \cite{LSC}.  

As a further check of the validity of $x_c$ in the context of the magnetic 
scales discussed here we performed ZF- $\mu $SR measurements for 
La$_{2-x}$Sr$_x$Cu$_{1-y}$Zn$_y$O$_4$ with $x=0.15$, 0.20, and
$y=0.01$, $0.02$. Substitution with Zn
slows down the spin fluctuations and greatly enhances the muon
depolarisation rate at low temperatures \cite{CP3,Julien,Nachumi1}. As a consequence
of this, both $T_g$ and $T_f$, are significantly enhanced. For $x(Sr)=0.15$ and 
$y=0.01$ and 0.02, we find that $T_g$ $=2.5K$ and $3.5K$ and $T_f$ $=15K$
and $25K$, respectively. However, for $x=0.20$ the muon depolarization rate
is unchanged by the Zn substitution. This result
also suggests that there is indeed a critical point in the slightly overdoped
region. Furthermore, the enhancement of the depolarisation rate by Zn doping confirms
that it is the spin fluctuations on the CuO$_2$ planes that are probed by $\mu $SR.

In conclusion, our results confirm that spin glass freezing is present 
in the superconducting state of underdoped LSCO \cite{Weid,Kiefl,CN}, 
but show more clearly that it extends slightly above optimal doping. 
We have also provided evidence for very slow 
spin fluctuations in the same doping region.
The doping level at which these characteristic features disappear is very close 
to the special point $x=0.19$ \cite{TallonLoram} suggesting a connection between 
the magnetic energy scales identified in this work, the normal state energy gap 
and the anomalous properties of underdoped HTS. Our comparison with the measured 
superfluid density implies that $low$ $frequency$ spin 
fluctuations compete with superconductivity in the cuprates. 
The special doping level $x=0.19$ could therefore mark the point at which 
these effects disappear and the remaining high frequency magnetic excitations 
are favourable for superconductivity in the cuprates. The competition
between quasi-static magnetic correlations with superconductivity 
is reminiscent of the way the Neel temperature 
of heavy fermion compounds goes to zero at a quantum critical point where the 
magnetic excitations are then most favourable for superconductivity.

We are grateful to A.D. Taylor of the ISIS Facility, Rutherford Appleton
Laboratory for the allocation of muon beam time and W.Y. Liang for
encouragement and support. C.P. thanks C. Bernhard, J.I. Budnick, S.
Chakravarty, S.M. Hayden, S. A. Kivelson, 
W.Y. Liang, P.B. Littlewood, J.W. Loram, Ch.
Niedermayer and C.M. Varma for useful discussions, and The Royal Society and 
Trinity College, Cambridge for financial support.

\end{multicols}


\begin{references}
\bibitem{SC}  S. Chakravarty, B.I. Halperin, and D.R. Nelson, Phys. Rev.
Lett. {\bf 60}, 1057 (1988).

\bibitem{CMV}  C.M. Varma $et$ $al.$, Phys. Rev. Lett. {\bf 63}, 1996 (1989).

\bibitem{SY}  S. Sachdev and J. Ye, Phys. Rev. Lett. {\bf 69}, 2411 (1992).

\bibitem{CC}  C. Castellani, C. DiCastro and M. Grilli, Phys. Rev. Lett. 
{\bf 75}, 4650 (1995).

\bibitem{CV1}  C.M. Varma, Phys. Rev. Lett. {\bf 83}, 3538 (1999).

\bibitem{SCRL}  S. Chakravarty, R.B. Laughlin, D.K. Morr and C. Nayak,
Phys. Rev. B {\bf 63}, 10000 (2001).

\bibitem{DP}  D. Pines, cond-mat/0002281.

\bibitem{Harshman} D. R. Harshman {\it et al.}, Phys. Rev. B {\bf 38}, 
852 (1988). 

\bibitem{Weid}  A. Weidinger $et$ $al.$, Phys. Rev. Lett. {\bf 62}, 102
(1989); 

\bibitem{Kiefl}  R.F. Kiefl $et$ $al.$, Phys. Rev. Lett. {\bf 63}, 2136
(1989).

\bibitem{Budnick} J. I. Budnick {\it et al.}, Europhys. Lett. 
{\bf 5}, 65 (1988); F. C. Chou {\it et al.}, Phys. Rev. Lett. 
{\bf 75}, 2204 (1995).

\bibitem{Berthier}  C. Berthier, M.-H. Julien, M. Horvatic and Y. Berthier,
J. Phys. I France {\bf 6}, 2205 (1996).

\bibitem{GA1}  S. M. Hayden {\it et al.}, Phys. Rev. Lett. {\bf 66}, 
821 (1991); S-W. Cheong $et$ $al.$, {\it ibid} 
{\bf 67}, 1791 (1991); T.E. Mason $et$ $al.$, {\it ibid} {\bf 71}, 919
(1993); H.A. Mook, $et$ $al.$, Nature (London) {\bf 395}, 580 (1998); 
B. Lake $et$ $al.$, {\it ibid} {\bf 400}, 43
(1999); S. Wakimoto $et$
$al.$, Phys. Rev. B {\bf 60}, R769 (1999); Y.S. Lee $et$ $al.$, {\it ibid} 
{\bf 60}, 3643 (1999).

\bibitem{KY}  K. Yamada $et$ $al.$, Phys. Rev. B {\bf 57}, 6165 (1998).

\bibitem{CN}  Ch. Niedermayer $et$ $al.$, Phys. Rev. Lett. {\bf 80}, 3843
(1998).

\bibitem{HA}  M. Havilio and A. Auerbach, Phys. Rev. Lett. {\bf 83}, 4848
(1999).

\bibitem{Mathur} N. Mathur $et$ $al.$, Nature (London) {\bf 394}, 39 (1998).

\bibitem{JWL}  J.W. Loram, K.A. Mirza, J.R. Cooper and J.L Tallon, J. Phys.
Chem. Solids {\bf 59}, 2091 (1998).

\bibitem{LSC}  C. Panagopoulos $et$ $al.$, Phys. Rev. B {\bf 60}, 14617
(1999).

\bibitem{TallonLoram}  J.L. Tallon and J.W. Loram, cond-mat/0005063.

\bibitem{YU}  Y.J. Uemura $et$ $al.$, Phys. Rev. B {\bf 31}, 546 (1985).

\bibitem{Torrance}  J.B. Torrance $et$ $al.$, Phys. Rev. Lett. {\bf 61}, 1127 (1988).

\bibitem{EL}  D. Lampakis $et$ $al.$, Phys. Rev. B {\bf 62}, 8811 (2000).

\bibitem{JRC}  J.R. Cooper and J.W. Loram, J. Phys. I France {\bf 6}, 2237
(1996).

\bibitem{Loram}  J.W. Loram $et$ $al.$, Proceedings of the 10th Ann. HTS
Workshop (World Scientific, Singapore 1996) p. 341.

\bibitem{CP4}  C. Panagopoulos $et$ $al.$, Phys. Rev. B {\bf 61}, R3808 (2000).

\bibitem{PGR}  P.G. Radaelli $et$ $al.$, Phys. Rev. B {\bf 49}, 4163 (1994).

\bibitem{muonsite}  B. Nachumi $et$ $al.$, Phys. Rev. B {\bf 58}, 8760 (1998).

\bibitem{CP1}  C. Panagopoulos $et$ $al.$, Phys. Rev. Lett. {\bf 79}, 2320
(1997).

\bibitem{CP2}  Y.J. Uemura $et$ $al.$, Phys. Rev. Lett. {\bf 62}, 2317 (1989).

\bibitem{CP3}  We find that the depolarisation rate is systematically
enhanced at low temperatures, typically below $0.3T_c$, with decreasing $x$
(for $x\leq 0.15$). In particular for $x<0.10$ the enhancement in the muon
spin depolarisation is more than $100\%$ of the the value above $0.3T_c$. An
example is given in Fig. 1(a) in: C. Panagopoulos $et$ $al.$, Phys. Rev. B 
{\bf 60}, 14617 (1999).

\bibitem{BDR}  R. Cywinski and B.D. Rainford, Hyperfine Interact. {\bf 85},
215 (1994).

\bibitem{Kiv}  V.J. Emery and S.A. Kivelson, 
J. Phys. Chem. Solids {\bf59}, 1705 (1998); J. Zaanen, {\it ibid}
{\bf59}, 1769 (1998). 

\bibitem{Julien}  M.-H. Julien $et$ $al.$, Phys. Rev. Lett. {\bf 84}, 3422
(2000).

\bibitem{Nachumi1}  B. Nachumi $et$ $al.$, Phys. Rev. Lett. {\bf 77}, 5421
(1996); B. Nachumi $et$ $al.$, Hyperfine Interact. {\bf 105},
125 (1997).


\end{references}
\end{document}